\documentclass[a4paper]{article}

\usepackage{graphicx}
\usepackage[latin1]{inputenc}
\usepackage{amsmath,amsthm}
\usepackage[OT1]{fontenc}
\usepackage{parskip} \parskip=1ex plus 5pt minus 1pt \parindent=4ex
\usepackage{RR}

\usepackage{mymacros}
\usepackage[obeyspaces]{url} 

\RRauthor{
	 Marc Shapiro \and Nuno Preguiça
}
\authorhead{Shapiro, Preguiça}
\RRdate{Octobre 2007}
\RRtitle{Conception d'un type de données répliqué commutatif}
\titlehead{Designing a commutative replicated data type}
\RRetitle{
   Designing a commutative replicated data type%
\thanks{
    This work is supported in part by the EU FP6 project Grid4All, the
    French ARA project Respire, the French ARC project Recall, and the
    Portuguese FCT/MCTES project POSC/59064/2004, with FEDER funding.
}
}
\RRabstract{
    Commuting operations greatly simplify consistency in distributed
    systems.
    This paper focuses on designing for commutativity, a topic neglected
    previously.
    We show that the replicas of \emph{any} data type for which
    concurrent operations commute converges to a correct value, under
    some simple and standard assumptions.
    We also show that such a data type supports transactions with very
    low cost.
    We identify a number of approaches and techniques to ensure
    commutativity.
    We re-use some existing ideas (non-destructive updates coupled with
    invariant identification), but propose a much more
    efficient implementation.
    Furthermore, we propose a new technique, background
    consensus.
    We illustrate these ideas with a shared edit buffer data type.
}
\RRresume{
    La commutativité des opérations simplifie grandement la cohérence
    dans les systèmes répartis.
    Ce papier aborde la conception visant la commutativité, qui est un
    sujet négligé.
    Nous démontrons que les réplicats tout \emph{tout} type de données,
    dont les opérations concurrent commutent, convergent vers une valeur
    correcte, sous des hypothèses simples et courantes.
    Nous montrons aussi qu'un tel type de données peut exécuter des
    transactions à un coût très faible.
    Nous identifions quelques approches et quelques techniques qui
    assurent la commutativité.
    Nous réutilisons quelques idées existantes (les mises à jour non
    destructives couplées à une identification invariante) mais nous en
    proposons une réalisation beaucoup plus efficace qu'auparavant.
    De plus nous proposons une nouvelle technique, celle du consensus en
    tâche de fond.
    Nous illustrons ces idées sur un example de tampon d'édition
    partagé.
}
\RRmotcle{Réplication des données, réplication optimiste, opérations
commutatives}
\RRkeyword{Data replication, optimistic replication, commutative
operations}
\RRprojet{Regal}
\RRtheme{\THCom}
\URRocq

\begin{document}
\makeRR

\section{Introduction}
\label{sec:introduction}

To share information, users located at several sites may concurrently
update a common object, e.g., a text document.
Each user operates on a separate \emph{replica} (i.e., local copy) of the
document.
A well-studied example is co-operatively editing a shared text.

As users make modifications, replicas diverge from one another.
Operations initiated on some site propagate to other sites and are
\emph{replayed} there.
Eventually every site executes every action.
Even so, if sites execute them in different orders, their replicas
might still not converge.
Various solutions are available in the literature; for instance,
serialising the actions \cite{con:rep:615} or operational transformation
\cite{syn:app:1430}.
Such designs are usually complex and non-scalable; thus, despite an
extensive literature, there is still no satisfactory solution to the
shared text editing problem.

We suggest a different approach: design replicated data types such that
operations commute with one another.
Let us call such a type a \emph{commutative replicated data type} or
CRDT\@.
CRDT replicas provably converge.
Furthermore, CRDTs support transactions ``for free.''
However, designing a non-trivial CRDT is difficult.

Although the advantages of commutativity are well known, the problem of
designing data types for commutativity has been neglected.
Recently, Oster et al.\ proposed a replicated character buffer CRDT
called WOOT \cite{app:rep:1587}.
WOOT operations commute, because updates are non-destructive, and because
the the identity of a character does not change with concurrent edits.
However, WOOT has some drawbacks: it wastes a lot of space, and it does
not support block operations such as cut-and-paste.

This paper presents the design of a non-trivial CRDT for concurrent
editing, called \emph{treedoc}.
Since it is a CRDT, convergence is guaranteed.
It supports block operations.
Space overhead is kept to a minimum: there is no to little internal
meta-data; deleted information can be forgotten; and identifiers are
kept short.
Common edit operations respond locally and suffer no network latency.
Treedoc is fault-tolerant and supports disconnected operation.

As in WOOT, ordinary editing operations are non-destructive and
identification does not change with concurrent edits, but our
implementation is very different from WOOT.
The basic treedoc structure is a binary tree of atoms.
The path to a node is a bitstring.
For efficiency, structural operations switch between a
flat buffer and a tree.
These operations are potentially non-commutative; to avoid this problem,
structure changes rely either on common knowledge or on consensus.
To avoid the latency associated with consensus, it occurs in the
background (not in the critical path of editing operations) and aborts
if it conflicts with an edit.

In summary, the contributions of this paper are the following:
  \begin{itemize}
    \item
          A design principle: concurrent operations should commute.
          We prove that \emph{any} Commutative Replicated Data Type (CRDT)
          converge, under some simple and standard assumptions.
    \item
          We identify two alternative approaches to commutativity:
          operation coalescing vs.~precedence.
          Coalescing is better, but is not always possible.
    \item
          The design of treedoc, a non-trivial, space-efficient,
          responsive, coalescing CRDT for distributed editing.
    \item
          We identify some previously-published techniques for
          coalescing, such as non-destructive
          update and invariant identity.
          We propose a novel implementation of these techniques.
    \item
          We propose a novel technique for coalescing: where a consensus is
          necessary, it is restricted to non-essential operations, occurs in the
          background, and aborts if it conflicts with an essential
          operation.
    \item
          We show that a CRDT readily supports transactions at
          a very low implementation cost.
  \end{itemize}

The paper proceeds as follows.
Section~\ref{sec:introduction} is this introduction.
We describe our system model in Section~\ref{sec:model}.
Section~\ref{sec:sharedbuffer} describes the shared buffer abstract data
type.
We suggest a simple implementation of this data type in
Section~\ref{sec:simple}.
In Section~\ref{sec:mapping}, we examine how to convert to a more
efficient representation and back.
Section~\ref{sec:mixed} explains
the full treedoc implementation, combining the advantages of the two
preceding sections.
We build transactions on top of a CRDT in
Section~\ref{sec:transactions}.
Section~\ref{sec:previous} compares with previous work.
Section~\ref{sec:conclusion} concludes.
We provide a proof of convergence in Appendix~\ref{apx:proof}.

\section{System model}
\label{sec:model}

\subsection{Replicated execution and eventual consistency}

We consider an asynchronous distributed system, consisting of $N$
\emph{sites} (computers) connected by a network.
Communication between connected sites is reliable.
A site may disconnect but eventually reconnects.
We assume an epidemic style of communication, i.e., a site connects
at arbitrary intervals with arbitrary other sites, sending both local
updates and those previously received from other sites.
Eventually, every update reaches every site, either directly or
indirectly.

With no loss of generality, we consider a single object replicated at
any number of sites.
A user accesses the object through his local replica, \emph{initiating}
operations at the current site.%
\footnote{
    We assume that a given operation is initiated at a unique site.
}
The operations execute locally and are logged.
Eventually the log is transmitted and the operations it contains are
\emph{replayed} at other sites.
Eventually all sites execute the same operations (either by local
submission or by remote replay), in some sequential order, but not
necessarily in the same order.

We say operation $o$ happens before $o'$ (noted $o \happensbefore o'$)
if some site initiates $o'$ after the same site has executed $o$.%
\footnote{
    The site executes $o$ either because it was initiated locally, or
    because it was initiated at another site and delivered here in a
    message.
    Thus the $\happensbefore$ relation is identical to Lamport's
    happens-before \cite{con:rep:615}.
}
We require that if $o \happensbefore o'$, then all sites execute $o$
before $o'$ (not necessarily immediately before).
Common epidemic protocols, such as Bayou's anti-entropy
\cite{rep:con:1414}, ensure this so-called ``causal ordering'' property.
To implement this property, it suffices to delay the execution of some
operation $o$, until all operations that happen before $o$ have been
executed.
Well-known techniques such as vector clocks or version vectors
\cite{alg:rep:738} can be used to track happens-before dependencies.

Operations are \emph{concurrent} if neither happens before the
other: $o \concurrent o' \equaldef \neg ( o \happensbefore o') \land
\neg ( o' \happensbefore o)$.

Two operations $o$ and $o'$ \emph{commute} iff, whatever the current
state of the object, such that execution of either $o$ or $o'$ succeeds,
executing $o$ immediately followed by $o'$ also succeeds, and leads to
the same state as executing $o'$ immediately followed by $o$.

A \emph{Commutative Replicated Data Type} (CRDT) is a data type where
all concurrent operations commute with one another.
We prove (in Appendix~\ref{apx:proof}) that CRDTs guarantee
\emph{eventual consistency}: provided that every site executes every
operation in an order consistent with happens-before, the final state of
replicas is identical at all sites.

Furthermore, CRDTs support serialisable transactions with virtually no
overhead.
If all operations commute, so do arbitrary sets of operations.
If every site executes transactions sequentially, in an order consistent
with happens-before, the local orders are all equivalent.
This ensures serialisability.
Furthermore, transactions never abort.
Hence, very little mechanism is needed.
We return to transactions in Section~\ref{sec:transactions}.

\subsection{Ensuring that operations commute}

Two operations \Alpha and \Beta commute if, for any state $T$, execution
sequences \sequence{T \seq \alpha \seq \beta} and \sequence{T \seq \beta
\seq \alpha} are both correct states and are equivalent.
There are two basic approaches for ensuring commutativity, which we call
coalescing and precedence.

Intuitively, coalescing means that \Alpha preserves the effect of \Beta
and vice-versa; i.e., the post-condition of both operations is
satisfied, whatever their relative execution order.%
\footnote{
    This is sometimes called ``intention preservation.''
}
This is the standard meaning in mathematics, for instance when we say
that addition and subtraction of integers commute.

The alternative is to define an order of precedence, say \Beta takes
precedence over \Alpha: when both operations execute in either order,
\Beta takes effect, but not necesariy \Alpha.
A typical implementation is that in the order \sequence{\alpha \seq \beta}, the
latter overwrites the results of the former, and in the order \sequence{\beta \seq
\alpha}, the latter is replaced by a no-op.
For instance, most replicated file systems follow the ``Last Writer
Wins'' rule \cite{rep:syn:1500}: when two users write to the same file,
the write with the highest timestamp takes precedence.
The write with the lowest timestamp may be lost.
(In contrast, writes to different files coalesce.)

Clearly, the coalescing approach is preferable to precedence; but
precedence is much easier to achieve.

\section{Shared buffer replicated data type}
\label{sec:sharedbuffer}

We consider a shared, replicated document, consisting of a linear
sequence of atoms.
An atom may be a character or some other immutable payload, e.g., a
graphical illustration inserted inside the document.
We designed treedoc it to be as unrestricted and flexible as possible,
to enable a variety of applications to use it.

Each user has a copy of the document.
Each user can modify his replica independently by executing two types of
\emph{edit operations}:
  \begin{itemize}
    \item
          \insertatom{\insertpos}{\newatom}{S} visibly inserts
          atom \newatom in the document.
          (In the underlying data structure, there may be other data before or
          after \insertpos, but it is not visible to the user.)
          All atoms at positions strictly less than $\insertpos$ lie to the left
          of \newatom; all those strictly greater than \insertpos lie to its
          right.
          The $S$ argument is the initiating site, as justified later.
    \item
          \deleteatom{\delpos}{S} visibly removes the atom existing at position
          \delpos.
          (The atom may still be in the data structure but is
          not visible to the user any more.)
          The $S$ argument is the initiating site.
  \end{itemize}

We defer to Section~\ref{sec:transactions} the description of a
transaction construct, enabling atomic bulk operations such as cutting
and pasting a block of text, or searching and replacing all instances of
a pattern.

Our treedoc design ensures commutativity by coalescence, i.e., the
effect of \insertatomop and \deleteatomop is the same at all sites.

At a higher level, the application might have stronger requirements.
For instance, it might disallow inserting characters inside deleted
text; or it might ensure a proper hierarchy of chapters, sections and
paragraphs; etc.
Enforcing such semantic conflicts requires higher-level conflict
detection and resolution mechanisms, which are non-commutative, but are
out of our focus.

Similarly, we emphasise that the stringtree structure is completely
decoupled from any higher-level document hierarchy (e.g., XML tree
structure).

\subsection{Unique identifiers for positions}
\label{sec:uid-crdt}

Let us assume the existence of unique position identifiers, with the
following properties:
    \begin{itemize}
      \item
            Each postion in the atom buffer has an identifier that is
            unique with respect to all other positions, and that remains
            constant, for the whole lifetime of the document.%
\footnote{
    However, an unused identifier can be garbage-collected and re-used.
    We do not attempt to formalise this property.
}
      \item
            There is a total order of position identifiers, noted $<$.
      \item
            Given any two identifiers $L$ and $R$ such that $L < R$, it
            is possible to generate a fresh unique identifier $N$ such
            that $L < N < R$.
    \end{itemize}

We will call these identifiers UIDs (unique identifiers).
Real numbers have the properties required for UIDs, but the third
property requires infinite precision, which is not realistic.
In Section~\ref{sec:simple} we will present a practical alternative,
based on trees.

\subsection{Abstract atom buffer CRDT}

Consider an abstract data type whose state $T$ is a set of
$(\uid, \atom)$ couples, where \uid{}s are unique.
The content of state $T$ is the sequence of all $\atom$s in $T$ ordered
by their $\uid$.
Operation \insertatom{u}{a}{S} adds the pair $(u, a)$ to the set.
If a pair $(u, a)$ exists in the set, operation \deleteatom{u}{S}
removes the pair, whatever $a$.
We now prove that concurrent operations of this data type commute.

\begin{lemma}
  Insert operations commute.
  For any data state $T$, any fresh unique identifiers $u_{1}$ and
  $u_{2}$, any atoms $a_{1}$ and $a_{2}$, and any originating sites
  $S_{1}$ and $S_{2}$: $\sequence{T \seq \insertatom{u_{1}}{a_{1}}{S_{1}}
  \seq \insertatom{u_{2}}{a_{2}}{S_{2}}} \equiv \sequence{T \seq
  \insertatom{u_{2}}{a_{2}}{S_{2}} \seq
  \insertatom{u_{1}}{a_{1}}{S_{1}}}$.
\end{lemma}

\begin{proof}
  After executing the two insert operations, the resulting state
  includes the two new atoms.
  Furthermore, atoms are ordered by unique identifiers.
  Therefore, the final state is the same.
\end{proof}

\begin{lemma}
  An insert operation commutes with a delete operation when they refer to
  different unique identifiers.
  For any state $T$, any fresh unique identifier $u_{1}$, any
  unique identifier $u_{2} \neq u_{1}$, any atom
  $a_{1}$, and any originating sites $S_{1}$ and $S_{2}$:
  $\sequence{T \seq \insertatom{u_{1}}{a_{1}}{S_{1}} \seq
  \deleteatom{u_{2}}{S_{2}}} \equiv \sequence{T \seq \deleteatom{u_{2}}{S_{2}} \seq
  \insertatom{u_{1}}{a_{1}}{S_{1}}}$.
\end{lemma}

\begin{proof}
  Two cases must be considered.
%
  First, when $T$ includes the atom with identifier $u_{2}$.
  By executing both operations in any order, the final state of $T$ will
  include an additional atom identified by $u_{1}$ and it will not
  include the atom identified by $u_{2}$.
  As atoms are ordered by their unique identifier, the final state is
  the same.
  Second, when $T$ does not include the atom with identifier
  $u_{2}$.
  By executing both operations in any order, the final state of $T$ will
  include an additional atom identified by $u_{1}$ (the atom
  identified by $u_{2}$ was not present in the original state.
  As atoms are ordered by their unique identifier, the final state is
  the same.
\end{proof}

\begin{lemma}
  If an insert operation and a delete operation refer to the same
  unique identifier, then the insert happens-before the delete.
\end{lemma}

\begin{proof}
  According to the specification of Section~\ref{sec:sharedbuffer},
  a user may initiate operation \deleteatom{u}{S} at site $S$ only if a
  pair $(u,a)$ exists (for some $a$) in the current state at site $S$.
  This pair must have been inserted by an \insertatomop operation
  executed previously at site $S$.
\end{proof}

\begin{lemma}
  Delete operations commute.
  For any state $T$, any unique identifiers $u_{1}$ and $u_{2}$ and any
  originating sites $S_{1}$ and $S_{2}$:
  $\sequence{T \seq
  \deleteatom{u_{1}}{S_{1}} \seq \deleteatom{u_{2}}{S_{2}}} \equiv \sequence{T \seq
  \deleteatom{u_{2}}{S_{2}} \seq \deleteatom{u_{1}}{S_{1}}}$.
\end{lemma}

\begin{proof}
  For any original state $T$, the final state will not include the atoms
  identified by $u_{1}$ and $u_{2}$, but it will include all
  other atoms, as no other atom will ever has the same unique
  identifier.
  Thus, the final state will include the same set of atoms and, as atoms
  are ordered by their unique identifier, the final state is exactly the
  same.
\end{proof}

    \begin{theorem}
      The data type described in this section is a CRDT\@.
    \end{theorem}
    \begin{proof}
      By the above lemmas, all concurrent operation pairs (insert-insert,
      delete-delete, insert-delete) commute.
    \end{proof}

\section{Treedoc abstract data type}
\label{sec:simple}

We start with a very simple design, which satisfies the coalescence
requirement
, but has some limitations.
In later sections, we will improve the design.

\subsection{Paths}

We manage the document as a binary tree.
A tree node contains either a single atom, or nil.
The identifier of an atom is its \emph{path} in the tree.
The path to the root is the empty bitstring \emptypath; the path
concatenation operator is noted $\concat$.
The left child of a node is 0, its right child is 1.
We walk the tree in infix order, skipping nil nodes (but not their
descendants).

\begin{figure}[h]
  \begin{center}
    \includegraphics[height=3cm]{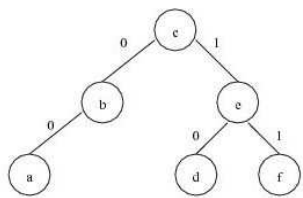}
  \end{center}
  \caption{Identifiers in a shared text buffer}
  \label{fig:simple}
\end{figure}

For example, Figure~\ref{fig:simple} represents the document state
\charstring{abcdef}, with the following identifiers:
$\id{\chars{a}}=[00]; \id{\chars{b}}=[0];
\id{\chars{c}}=[]; \id{\chars{d}}=[10]; \id{\chars{e}}=[1];
\id{\chars{f}}=[11]$.

We define the following partial order over identifiers.
Node $id_1$ is to the left of $id_2$ (or, equivalently, $id_2$ is to the
right of $id_1$), noted $id_1 < id_2$, iff:
\begin{itemize}
  \item
        $id_1=[c_1...c_n]$ is a prefix of $id_2=[c_1 ... c_n j_1 ...
        j_m]$ and $j_1 = 1$, or
  \item
        $id_2=[c_1...c_n]$ is a prefix of $id_1=[c_1 ... c_n i_1 ...
        i_m]$ and $i_1 = 0$, or
  \item
        $id_1=[c_1...c_n i_1...i_n]$ has a common prefix with
        $id_2=[c_1...c_n j_1...j_m]$ and 
        $i_1 =
        0$. The prefix may be empty.
\end{itemize}

We also define the ancestry of a node.
Node $u$ is the (direct) parent of node $v$, noted \parent{u}{v}, iff
$\id{v} = \id{u} \concat 0 \lor \id{v} = \id{u} \concat 1$;
equivalently, $v$ is a (direct) child of $u$.
Node $u$ is an ancestor of $v$ (or, equivalently, $v$ is a descendant of
$u$), noted \ancestor{u}{v}, if $u$ is a parent, or grand-parent,
or great-grand-parent, etc., of $v$.

\subsection{Deleting}

We start with the simplest procedure, deleting an atom: simply replace
the content of the node with nil.
Since the identification of the deleted node is unique, it is clear that
the initiator and replay executions will all delete the same node.
Sometimes, during replay, the node to be deleted may not exist, but this
can only be because it was already deleted previously.

We will say that the delete is \emph{stable} once it has been executed
at all nodes.
No operation that happens-after the delete is stable will ever refer to the
node identifier; therefore, if the node is a leaf, it can be completely
forgotten (and so on recursively).
Thus a subtree that contains only stably deleted nodes can be completely
removed and forgotten.

To this effect we introduce a \gc{N} procedure that removes leaf $N$ if
it is stably deleted.
A node may call \gc{N} at any time after $N$ is deleted.
Operation \gcop is local only, it does not have a replay version.

Procedure \stabledel{N} tests for stability.
Conceptually, \stabledel{N} waits for acknowledgments from all sites that
have executed \delete{N}.
We refer to Golding \cite{pan:pro:813} for an efficient implementation
of stability that compacts acknowledgments for all past operations into a
single vector clock or matrix clock.

\subsection{Implementing inserts}

To insert \newatom between $\atom_p$ and $\atom_f$, we must grow the
tree in a way that satisfies the relation $\id{\atom_p} < \id{\newatom}
< \id{\atom_f}$.
In this section, we present a very simple algorithm that does not attempt
to balance the tree; later, we will resolve this issue.

\begin{algorithm}[tbp]
  \caption{New unique identifier for insert}
  \label{alg:simple-insert-treedoc}
  \begin{algorithmic}[1]
    \Function{\newuid}{$(\atom_p,\uid_p),(\atom_f,\uid_f)$}
        \Comment{$(\atom_p,\uid_p)$: previous atom; $(\atom_f,\uid_f)$: following atom;}
        \Require $\uid_p < \uid_f$
        \If{$\exists (\atom_m,\uid_m): \uid_p < \uid_m < \uid_f$}
        \Return $\newuid( (\atom_p,\uid_p), (\atom_m,\uid_m) )$
       \ElsIf{$\ancestor{(\atom_p,\uid_p)}{(\atom_f,\uid_f)}$}
        \Return{$\uid_f \concat 0$}
        \ElsIf{$\ancestor{(\atom_f,\uid_f)}{(\atom_p,\uid_p)}$}
        \Return{$\uid_p \concat 1$}
        \Else~
        \Return{$\uid_p \concat 1$}
        \EndIf
     \EndFunction
  \end{algorithmic}
\end{algorithm}

Algorithm~\ref{alg:simple-insert-treedoc} starts by checking whether
there is a node between $\atom_p$ and $\atom_f$.
If so, it recursively looks for the leftmost predecessor of $\uid_{f}$
that remains to the right of $\uid_{p}$.
When there is no node between $\atom_p$ and $\atom_f$, three cases may occur:

\begin{itemize}

  \item Node $\uid_p$ is an ancestor of $\uid_f$, i.e., $\uid_f$ is a
        right descendant of $\uid_p$.
        In this case, $\uid_f$ has no left child, so we create a new
        left child of node $\uid_f$.
        The new identifier is  $\uid_f \concat 0$. 

  \item
        Or, symmetrically, node $\uid_f$ is an ancestor of $\uid_p$.
        The new identifier us $\uid_p \concat 1$.

  \item
        Or, neither is an ancestor of the other.
        In this case, $\uid_p$ has no right child, so we create a new
        right child of node identified $\uid_p \concat 1$. 
\end{itemize}

\begin{figure}[h]
  \begin{center}
    \includegraphics[height=4cm]{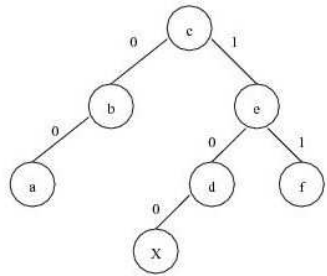}
  \end{center}
  \caption{Identifiers after inserting a new character atom}
  \label{fig:simple:insert}
\end{figure}

In the example of Figure~\ref{fig:simple}, for inserting
atom \chars{X} between \chars{c} and \chars{d}, a left child is created
under \chars{d} with identifier $[100]$, as shown in
Figure~\ref{fig:simple:insert}.

\subsection{Concurrent inserts}

In case of concurrent updates, a binary tree becomes insufficient,
because two users can concurrently insert an atom at the same position.
We maintain the basic binary tree structure, but we extend a node to
contain any number of \emph{side nodes} (and their descendence),
disambiguated by a $(\siteID,\counter)$ pair, where \siteID identifies
the initiator site.
We assume a total order of site identifiers, hence of disambiguators,
hence of side nodes: $(s_{1},c_{1}) < (s_{2},c_{2})$, iff $c_{1} <
c_{2}$ or $c_{1} = c_{2}$ and $s_{1} < s_{2}$.

\begin{figure}[h]
  \begin{center}
    \includegraphics[height=4cm]{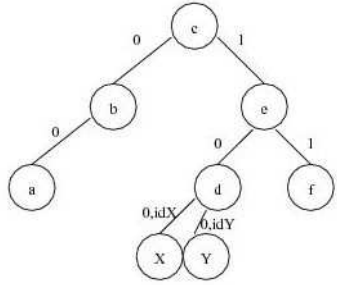}
  \end{center}
\caption{Side nodes and their identifiers, after concurrent inserts
         at position $[100]$} \label{fig:simple:doubleinsert}
\end{figure}

Algorithm~\ref{alg:simple-insert-treedoc} generates new unique
identifiers for insertion.
Figure~\ref{fig:simple:doubleinsert} shows an example of the situation.
Assuming that characters \chars{X} and \chars{Y} were inserted with the associated disambiguator \idX
and \idY respectively, we have $\id{\chars{X}}=[(1)(0)(0,\idX)]$ and
$\id{\chars{Y}}=[(1)(0)(0,\idY)]$.

\begin{figure}[h]
  \begin{center}
    \includegraphics[height=4cm]{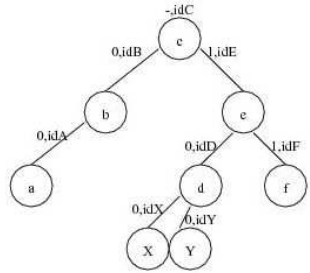}
  \end{center}
  \caption{Fully expanded identifiers, after concurrent inserts
           at position $[100]$}
  \label{fig:simple:doubleinsertcomplete}
\end{figure}

Since a concurrent update can occur at every level, conceptually, every
node may include a disambiguator.
Thus, in the example, we could
have, for example, $\id{\chars{d}}=[(--,\idC)(1,\idE)(0,\idD)]$.
The full state is as in Figure~\ref{fig:simple:doubleinsertcomplete}.

When disambiguators are used, the total order among identifiers is
defined as follows: Node $id_1$ is left of node $id_2$ (or,
equivalently, $id_2$ right of $id_1$), noted $id_1 < id_2$, iff:
\begin{itemize}
  \item
        $id_1=[c_1...c_n]$ is a prefix of $id_2=[c_1 ... c_n j_1 ...
        j_m]$ and $j_1 = (1,*)$, or
  \item
        $id_2=[c_1...c_n]$ is a prefix of $id_1=[c_1 ... c_n i_1 ...
        i_m]$ and $i_1 = (0,*)$, or
  \item
        $id_1=[c_1...c_n i_1 ... i_n]$ has a common prefix with
        $id_2=[c_1 ... c_n j_1 ... j_m]$ and $i_1=(0,*) \wedge j_1=(1,*)$ or $i1=(k,d_1) \wedge j1=(k,d_2) \wedge d_1 < d2$. The prefix may be empty.
\end{itemize}

To generate a new unique identifier, the algorithm~\ref{alg:simple-insert-treedoc} is used, with the returned identifier extended with a freshly generated disambiguator.

Once all concurrent inserts at the same location have executed at some
site, redundant disambiguators can be removed.
We will say that an insert is stable at some site, once that site has
received from all other sites some operation that happens-after the
insert.
At that point, it is guaranteed not to receive another concurrent insert
at the same node.
(To ensure this happens quickly, sites that are not actively editing
should send out occasional no-ops.)
Procedure \cleansideop removes redundant disambiguators; it is a local
procedure (it has no replay version).


\subsection{Treedoc abstract data type}
\label{sec:specification-simple}

Algorithm~\ref{alg:simple-treedoc} contains detailed specification of
the simple treedoc data type.
In addition to the operations given at the beginning of
Section~\ref{sec:sharedbuffer}, we specify \gcop and \cleansideop as
explained above.

The initiator versions of \insertatomop and \deleteatomop have pre-conditions,
to make sure that the user only addresses valid nodes, and to avoid
wastage of space.
However, there may be no restrictions on the replay version.
Therefore, the replay version has no precondition, and simply re-creates
any nodes that it may be missing.

This data type satisfies the coalescence requirement.
Since every atom has an identifier that does not change with other
operations, replaying a delete removes the intended atom.
Inserting an operation with a path positions it with respect to its left
and right neighbours, replaying an insert preserves the intended
location.
Since this data type is a CRDT, replicas are guaranteed to converge to
the same (correct) value.


In Algorithm~\ref{alg:simple-treedoc}, the notation \side{N}{\siteID}
stands for the side node of $N$ identified by $\siteID$.
The notation \sideOrNode{N}{\siteID} stands for \side{N}{\siteID},
if it exists, and $N$ otherwise.
\IamInitiator is true on the initiator site, and false on all replay
sites.

\begin{algorithm}[tbp]
  \caption{Simple treedoc}
  \label{alg:simple-treedoc}
  \begin{algorithmic}[1]
    \Procedure{\contentsop}{$N$}
        \Comment{$N$: a treedoc node}
        \State Walk subtree rooted at $N$ in infix order
        \State Return the atoms of the non-empty nodes
      \EndProcedure

\Statex

    \Procedure{\insertatomop}{${A},{N},{S}$}
        \State \Comment{$A$: atom to insert}
        \State \Comment{$N$: insertion position, chosen by \newuid}
        \Require $S$ = the initiator site
        \Require $A \neq \nil$
        \Require $\IamInitiator \implies$ all ancestors of node $N$ exist
        \Require $\IamInitiator \implies$ \sideOrNode{N}{S} does not exist
        \State Create any missing ancestors of $N$
        \State If necessary, create node $N$
        \State Create side node $n = \side{N}{\siteID}$ with $n.\contains = A$
     \EndProcedure
     \Statex
     \Procedure{\deleteatomop}{${N},{S}$}
        \Comment{$N:$ node to be deleted}
        \Require $\IamInitiator \implies \sideOrNode{N}{S}$ exists and
                  $\sideOrNode{N}{S}.\contains \neq \nil$
        \Require $S$ = the initiator site
        \State Create any missing ancestors of $N$
        \State If necessary, create node $N$
        \State $\sideOrNode{N}{S}.\contains \assign \nil$
        \State Send acknowledgment of \deleteatom{N}{S} to all sites
     \EndProcedure


\Statex

    \Procedure{\stabledelop}{$N$}
      \Comment{Await stable delete of node $N$}
      \If{acknowledgment for \deleteatom{N}{*} received from all sites}
          \Return \true
      \Else
          ~\Return \false
      \EndIf
    \EndProcedure

\Statex

    \Procedure{\gcop}{$N$}
      \Comment{$N$: a treedoc leaf}
      \Require{$N.\contents = \nil$}
      \Require{\Call{\stabledelop}{N}}
      \Require{\IamInitiator}
      \State Remove $N$
    \EndProcedure

\Statex

    \Procedure{\cleansideop}{$N$}
      \Comment{$N$: a treedoc node}
      \Require{\Call{\stableinsertop}{$N$}}
      \Require{\IamInitiator}
      \If {$| \{ \side{N}{S} | \forall S \} | = 1 $} \Comment{There is a
      single side node}
        \State $N \assign \side{N}{S}$  \Comment{Remove redundant
        disambiguator}
      \EndIf
    \EndProcedure

\Statex

    \Procedure{\stableinsertop}{$N$}
      \If {current site has received some operation that happens-after
         \insertatom{*}{N}{*} from every site}
         \State \Return \true
      \Else
         ~\Return \false
      \EndIf
    \EndProcedure

  \end{algorithmic}
\end{algorithm}

\section{Identifying a sequential buffer to a binary tree}
\label{sec:mapping}

The approach so far has a number of limitations.
Paths are variable length and can become very inefficient if the tree is
unbalanced (e.g., if users always append to the end of the buffer).
Concurrent inserts complicate the structure and the paths.
The tree metadata consumes memory; for instance, if atoms are one-byte
characters the overhead can be several times the payload.
Finally, deleted atoms that cannot be garbage-collected waste space.

Rather than attempt to fix each of these issues individually, we propose
a more radical solution.
In this section, we discuss structural operations that switch between
the efficient flat buffer representation, and the edit-oriented tree
representation.
The specification of these operations is as follows.
  \begin{itemize}
    \item
          \explode{\atomstring}.
          Returns a treedoc whose contents is identical to \atomstring.
    \item
          \flatten{\path}
          Returns an atom string whose contents is identical to the
          sub-treedoc rooted at \path.
  \end{itemize}

The initiator and replay versions of these operations must have
identical effect.
In particular, \explodeop must return exactly the same structure at all
sites.

Different implementations of \explodeop are possible, as long as it has
the same effect at every site.
Observing that the capacity of a complete binary tree with \depth levels
is $2^{\depth}-1$, we suggest the simple implementation in
Algorithm~\ref{alg:explode}.

  \begin{algorithm}[tbp]
    \caption{\explodeop and \flattenop}
    \label{alg:explode}
    \begin{algorithmic}[1]
      \Procedure{\explodeop}{\atomstring}
          \State $\depth = \lceil \log_{2} (\length{\atomstring} + 1) \rceil$
          \State $T$ = Allocate a complete binary tree of depth \depth
          \State Populate $T$ in infix order with the atoms of \atomstring
          \State Remove any remaining nodes
          \State Return $T$
      \EndProcedure
\Statex
      \Procedure{\flattenop}{N}
          \Comment{$N$: root of a subtree to be flattened}
          \State Walk subtree in infix order
          \State Return a linear buffer containing the atoms of the non-empty nodes
      \EndProcedure
    \end{algorithmic}
  \end{algorithm}

With these two operations, we can choose the string representation
or the treedoc representation.
The former is compact and efficient, but it does not readily support
concurrent edits.
When a treedoc becomes unbalanced or contains many nil nodes, it
suffices to \flattenop then \explodeop it to fix the problem.

However, these structural operations do not commute with edit
operations.
We study the solution of this problem next.

\section{Mixed tree}
\label{sec:mixed}

In this section, we study how to combine edit operations and structural
operations, while still retaining the advantages of a CRDT\@.

A first observation is that the \explodeop operation is not really
necessary.
Algorithm~\ref{alg:explode} can be interpreted as a mapping from a
string to a canonical treedoc representation.
Applying a path to a string implicitly converts the string to the
canonical treedoc.
Eliminating the explicit \explodeop operation removes the need to make
it commute with edits.

A second observation is that \flattenop is not an essential operation.
Aborting a \flattenop (leaving no side-effects) causes no harm.
Therefore, if \flattenop is concurrent with an edit operation in the
same subtree, we abort it.
More precisely, for \flattenop to take effect, it executes a distributed
commitment procedure.
When executing \flattenop at some site, if this site observes the
execution of a concurrent \insertatomop or \deleteatomop, that site
votes ``No'' to commitment, otherwise it votes ``Yes.''
The operation succeeds only if all sites vote ``Yes,'' otherwise it has
no effect.

Any distributed commitment protocol from the literature will do, for
instance two-phase commit or Gray and Lamport's
fault-tolerant protocol \cite{db:syn:1578}.

We may now envisage a mixed tree, where parts that are currently being
edited are in treedoc representation, and parts that are currently
quiescent are represented as strings.


\subsection{Fault tolerance and disconnected operation}

Ensuring fault tolerance and disconnected operation for disconnected
edits is straightforward.
Every site logs all its operations (whether locally initiated or remote)
on persistent storage.
When a site that was disconnected for some time reconnects with the rest
of the system, it simply exchanges with other sites the missing
information.
If a site fails and recovers the situation is the same.
If a site crashes, losing its memory, then when it restarts it behaves
like a new site, and copies over the state of some other site;
operations that it initiated before the crash and never sent to another
site are lost.

The situation is more complex for \flattenop{}s, since they require a
consensus.
To ensure that consensus is solvable in the presence of crashes, we
assume the existence of fault detectors \cite{rep:alg:1597}.

To allow disconnected operation, fault detectors must be capable of
distinguishing disconnection from a crash.
During the commit phase of \flattenop, a disconnected site is assumed to
be voting No, and \flattenop aborts.
This is distinct from a crashed site, which does not participate in the
commitment.

Similarly, \stabledelop should be modified to return true if all
non-crashed sites have acknowledged, and \stableinsertop should return
true if all non-crashed sites have sent an operation that happens-after
the insert.

Note that if a disconnected site is falsely diagnosed as crashed, any
operations that it initiated within a sub-tree that was \flattenop{}ed
cannot be replayed, because they use now-forgotten node identities.
Such operations are lost.
Similarly, if this site initiated operations that depend on a node
that was deleted and garbage-collected, then these operations are lost.

\section{Block edits and transactions}
\label{sec:transactions}

In practice, single-character edit operations are insufficient for
concurrent editing.
Users working on the same portion of text may find it unpleasant to see
their edits mixed together.
Furthermore, common operations such as as cutting and pasting, or global
replacement, are block operations.
We need transactions, to allow a user to insert, delete, replace or move
a block of text, without concurrent operations destroying the integrity
or location of the block.

Fortunately, a CRDT is ideal for building complex transactions out of
simple operations.
Since individual operations commute when concurrent, concurrent groups of
operations commute as well.
To ensure serialisability, it is sufficient to ensure that
transactions are executed sequentially (in any order compatible with
happens-before), whether at the initiator site or during replay.%
\footnote{
    For the purpose of sequential execution, an operation that is not
    part of any transaction is considered as a separate transaction of
    itself.
}

A transaction executes atomically (all-or-nothing), indivisibly (its
intermediate results cannot be observed) and durably (its results are
observable by all later operations).
We do not see any need for nested transactions.

Since individual operations commute, a transaction never aborts,
therefore transaction support can be very cheap.
All that is needed is some book-keeping of the beginning and end of
transactions, and buffering received operations to ensure sequential
execution.
While a site is executing a locally-initiated transaction, it buffers
remote operations, delaying their replay until the end of the
transaction.
When a site receives a remote transaction, it buffers the operations it
contains until the end of the transaction is received, and replays them
all at once.
Thus, \starttrans and \finishtrans are basically no-ops used
only as place-holders in the log.

  \begin{itemize}
    \item
          \starttrans opens a transaction.
          At the initiator site, the transaction will include all
          operations initiated at the same site, until the next
          \finishtrans.
          It is illegal to initiate two successive \starttrans
          operations without an intervening \finishtrans.
    \item
          \finishtrans closes the current transaction by the same
          initiator.
          It is illegal to initiate a \finishtrans unless a transaction
          is open.
  \end{itemize}
\noindent

Considering any two transactions (or isolated operations), either one
happens-before the other, or they are concurrent.
In particular, if a transaction contains an operation that edits node
$N$, and any operation of the same transaction is concurrent with
\flatten{N'}, and $\ancestor{N'}{N} \lor N = N'$, then the \flattenop
operation aborts.

Note that we could now define block operations such as
a block move or a global search-and-replace.
From the commutativity perspective, such new operation types are
considered equivalent to a transaction of \insertatomop and \deleteatomop
operations, but they can be implemented much more efficiently.

\section{Related work}
\label{sec:previous}

A comparison of several approaches to the problem of collaboratively
editing a shared text was written by Ignat et al.\ \cite{app:rep:1591}.

Operational transformation (OT) \cite{syn:app:1430} considers
collaborative editing based on non-commutative single-character
operations.
To this end, OT transforms the arguments of remote operations to take
into account the effects of concurrent executions.
OT requires two correctness conditions \cite{syn:app:1430}: the
transformation should enable concurrent operations to execute in either
order, and furthermore, transformation functions themselves must
commute.
The former is relatively easy.
The latter is more complex, and Oster et al.\ \cite{OsterRR05c} prove
that all existing transformations violate it.

OT attempts to make non-commuting operations commute after the fact.
We believe that a better approach is to design operations to commute in
the first place.
This is more elegant, and avoids the complexities of OT\@.

A number of papers study the advantages of commutativity for concurrency
and consistency control \cite[for instance]{syn:alg:1466,syn:1470}.
Systems such as Psync \cite{rep:alg:1594}, Generalized Paxos
\cite{rep:syn:1567}, Generic Broadcast \cite{rep:syn:1569} and IceCube
\cite{syn:1474} make use of commutativity information to relax
consistency or scheduling requirements.
However, these works do not address the issue of achieving
commutativity.

Weihl \cite{syn:1470} distinguishes between forward and backward
commutativity.
They differ only when operations fail their pre-condition.
In this work, we consider only operations that succeed at the submission
site, and ensure by design that they won't fail at replay sites.

Roh et al.\ \cite{app:rep:1588} were the first to suggest the CRDT
approach.
They give the example of an array with a slot assignment operation.
To make concurrent assignments commute, they propose a deterministic
procedure (based on vector clocks) whereby one takes precedence over the
other.

This is similar to the well-known Last-Writer Wins algorithm, used in
shared file systems.
Each file replica is timestamped with the time it was last written.
Timestamps are consistent with happens-before \cite{con:rep:615}.
When comparing two versions of the file, the one with the highest
timestamp takes precedence.
This is correct with respect to successive writes related by
happens-before, and constitutes a simple precedence rule for concurrent
writes.

In the precedence design of Roh et al., concurrent writes to the same
location are lost.
This is inherent to the destructive assignment operation that they
consider.
In ours, concurrent inserts are always coalesced, which is important in
order to support co-operative work.

In Lamport's replicated state machine approach \cite{con:rep:615}, every
replica executes the same operations in the same order.
This total order is computed either by a consensus algorithm such as
Paxos \cite{syn:rep:1548} or, equivalently, by using an atomic broadcast
mechanism \cite{alg:rep:1583}.
Such algorithms can tolerate faults.
However they are complex and scale poorly; consensus occurs within the
critical execution path, adding latency to every operation.

The precedence approach can be viewed as a poor-man's total order.
It does not require an online consensus algorithm, but it loses work.

In the treedoc design, common edit operations execute optimistically,
with no latency; it uses consensus in the background only.
Previously, Golding relied on background consensus for garbage
collection \cite{pan:pro:813}.
We are not aware of previous instances of background consensus for
structural operations, nor of aborting consensus when it conflicts with
essential operations.

\section{Conclusion}
\label{sec:conclusion}

It was known previously that commutativity simplifies consistency
maintenance, but the issue of designing systems for commutativity
was neglected.
This paper suggested a new paradigm for replication: the Commutative
Replicated Data Type or CRDT, designed such that concurrent operations
commute.
We prove that, under some simple and standard execution conditions,
replicas of any CRDT eventually converge.
This makes the implementation of replicated systems much simpler than
before.
Furthermore, CRDTs support transactions at very low cost.

However, designing a CRDT with the desirable property that no work is
lost (coalescence) is not easy.
We give a coalescing CRDT solution to the problem of a shared edit
buffer, by implementing some known techniques in a novel way (using
paths in a binary tree as invariant identifiers) and by some new
techniques (abortable consensus in the background).
This is possible only because updates are not destructive.

Our techniques are not limited to this particular problem, and are
generaliseable  non-destructive updates in other data structures, such
as directories.

We purposely designed treedoc to support arbitrary mixtures of edit
operations.
We separate out the issue of semantic constraints and conflict
detection, which we study elsewhere
\cite{app:rep:1547,syn:1474,rep:syn:1498,rep:syn:1482}.
We interpret conflicts as cases of irreducible non-commutativity.
Any real system must support a mix of data types, some coalescing, some
using precedence, and some not commutative.

Our next step in this research will be to enable peer-to-peer
co-operative editing at a large scale, by implementing treedoc within an
existing text editor or wiki system.
This will enable a deeper investigation of pragmatic issues and
performance studies.

\appendix

\section{Proof of eventual consistency}
\label{apx:proof}

We prove the following property.
    Assuming:
      \begin{itemize}
        \item
              That every operation, initiated at any site, eventually
              executes at all sites,
        \item
              That if $o \happensbefore o'$, then $o$ executes before $o'$ at
              every site,
        \item
              That all concurrent operations commute,
      \end{itemize}
    \noindent
    then the state of the object eventually converges at all sites.

Our proof is by recurrence over all legal execution schedules.

\subsection{Notation}

A multilog $M = (V, H)$ is a directed graph, consisting of a set of
operations $V = \{ \alpha, \beta, \ldots{} \}$ connected by the
happens-before relations $H = \{ (x, y) \in V \times V: x \happensbefore
y\}$.

A schedule $T = \sequence{\alpha \seq \beta \seq \ldots{}}$ of a multilog is a
sequential enumeration of operations in some order consistent with
happens-before.
Formally,
$
T = (L, <_{T}) \in \sched{(V, H)} \iff \Forall{x, y \in L}
          (x, y \in V)
    \land (x <_{T} y \implies x \neq y)
    \land (x \happensbefore y \implies x <_{T} y)
$.
The sequence operator is noted $ \seq $.

A schedule $T=(L,<_{T}) \in \sched{(V,H)}$ is said complete with respect
to its multilog, iff it contains all operations, i.e., iff $L = V$.

Operation $v$ extends schedule $T \in \sched{M}$ if \sequence{T \seq v} is in
\sched{M}.


A state \emph{quasi-state} is a schedule $T \in \sched{M}$ whose initial
element is distinguished operation \init (denoting the common initial
state).
We assume that we can further distinguish (in some application-specific
manner) between {illegal} and {legal} quasi-states.
Any extension of an illegal quasi-state is itself illegal.
A legal quasi-state will be called a \emph{state} henceforth: $T \in
\state{M}$.
We assume the existence of an equivalence relation between states
$\equiv$.
Like legality, equivalence is application dependent.

\subsection{Recurrence proof}

We require that all concurrent operations commute, i.e., given any state
$T$ and concurrent operations $x$ and $y$ that extend $T$, the sequences
\sequence{T \seq x \seq y} and \sequence{T \seq y \seq x} are states and are equivalent.
Formally:
  $
    \Forall{T \in \state{(V,H)}}
    \ForallSuchthat{x, y \in V}{x \concurrent y \land
                                \sequence{T \seq x} \in \state{(V,H)} \land
                                \sequence{T \seq y} \in \state{(V,H)}}
    \sequence{T \seq y \seq x} \in \state{(V,H)}
      \land \sequence{T \seq x \seq y} \equiv \sequence{T \seq y \seq x}
  $

Given a set of natural numbers $N = \{1, 2, \ldots{}, n-1 \}$, we note $\rho$
some permutation of $N$, with elements $\rho(1), \rho(2), \ldots{}, \rho({n-1})$.

  \begin{theorem}[All complete states of $M$ are equivalent]
    Let $M = (V,H)$ be a multilog of size $|V| = n$.
    Let $T = \sequence{\init \seq \alpha_{1} \seq \ldots{} \seq \alpha_{n-1}}$ be
    a complete state of $M$, i.e.,
      $      T \in \state{M}
       \land \{ \init, \alpha_{1}, \ldots{}, \alpha_{n-1}\} = V$.
    Let $\rho$ be some arbitrary permutation of
        $\{1, 2, \ldots{}, n-1\}$,
    and let ${T}_{\rho}$ denote the sequence
        \sequence{\init \seq \alpha_{\rho(1)} \seq \ldots{} \alpha_{\rho(n-1)}}.
    Then, if ${T}_{\rho}$ is a state, it is equivalent to $T$:
        ${T}_{\rho} \in \state{M} \implies T \equiv {T}_{\rho}$.
  \end{theorem}

The proof is by recurrence.
The theorem is obviously true for $n=1$ and $n=2$.
Assume it is true for abitrary $n$; we shall prove that it remains true
for $n+1$.
Let $T = \sequence{\init \seq \alpha_{1} \seq \ldots{} \seq \alpha_{n-1}}$ be a complete
state of $M=(V,H)$ where $n=|V|$.

Consider $M'=(V',H')$ such that $M \subseteq M' \land |V'| = n+1$.
$V$ and $V'$ differ by a single element, $\beta$.
With no loss of generality, we assume that $\beta$ does not happen
before any element of $V$, i.e., $\NotExists{x \in V}{(\beta,x)
\in H'}$.%
\footnote{
    In other words, the operations are sorted in
    $\happensbefore$ order before constructing the successive multilogs.
}
Note $T' = \sequence{T \seq \beta}$.

If $T' \notin \state{M'}$ the theorem is trivially true, because
$T'$ is not a state.
Therefore, assume $T' \in \state{M'}$.
It follows that $T'$ is a complete state of $M'$.

For some permutation $\rho$, let
    $T'_{\rho} = \sequence{\init \seq \alpha_{\rho(1)} \seq \ldots{} \seq \alpha_{\rho(n-1)} \seq \beta}$.
Consider now the set of sequences derived from $T'_{\rho}$
by permuting the position of $\beta$:
  \begin{center}
$$
  \begin{array}{c@{\,;\,}c@{\,;\,}c@{\,;\ldots;\,}c@{\,;\,}c@{\,;\,}c}
    \init & \alpha_{\rho(1)} & \alpha_{\rho(2)} & \alpha_{\rho(n-2)} & \alpha_{\rho(n-1)} & \beta \\
    \init & \alpha_{\rho(1)} & \alpha_{\rho(2)} & \alpha_{\rho(n-2)} & \beta             & \alpha_{\rho(n-1)} \\
    \init & \alpha_{\rho(1)} & \alpha_{\rho(2)} & \beta             & \alpha_{\rho(n-2)} & \alpha_{\rho(n-1)} \\
    \multicolumn{6}{c}{\ldots} \\
    \init & \alpha_{\rho(1)} & \beta           & \alpha_{\rho(n-3)} & \alpha_{\rho(n-2)} & \alpha_{\rho(n-1)} \\
    \init &  \beta          & \alpha_{\rho(1)} & \alpha_{\rho(n-3)} & \alpha_{\rho(n-2)} & \alpha_{\rho(n-1)} \\
  \end{array}
$$
  \end{center}

If the sequence on any line is not a state, then none of the following
lines is a state either; for these, the theorem is trivially true.

The first line is precisely $T'_{\rho}$.
If $T'_{\rho} \in \state{M'}$, then, by assumption,
   $T'_{rho} = \sequence{T_{\rho} \seq \beta} \equiv \sequence{T \seq \beta} = T'$.

Now examine the second line.
Either $\alpha_{\rho(n-1)} \concurrent \beta$, and they commute, and
therefore it is a state equivalent to the first line, and hence to $T'$;
or $\alpha_{\rho(n-1)} \happensbefore \beta$, and the second line is not a
state.
Similarly for all the following lines: each sequence is either not a state,
or is a state equivalent to $T'$.

Thus we have proven the recurrence clause for all permutations of
    $\{1, 2, \ldots{}, n\}$
that are in the same order as $\rho$.
Furthermore, since the recurrence clause is true for all permutations
$\rho$ of
    $\{1, 2, \ldots{}, n-1 \}$,
it is true for all permuations of $\{1, 2, \ldots{}, n\}$.
QED\@.

\subsection{Eventual consistency}

The above proves that, if different sites execute schedules consisting
of the same set of operations, the order of every schedule is consistent
with happens-before, and concurrent operations commute, then their final
states are equivalent.

If all clients stop initiating operations, and assuming that the system
transmits and executes every operation at all sites, then all replicas
converge to the same state.
This is the traditional definition of Eventual Consistency.
Our result is actually stronger, since the final state is a correct
state, and it includes all the submitted operations.

However, in a practical system, clients don't stop initiating
operations.
We can prove nonetheless that, for any time $t$, the state at every site
eventually includes a an equivalent prefix containing all operations up
to $t$.
Assume that initiating an operation is atomic.
Consider the set $O$ of operations initiated at all sites up to time
$t$, and the set $O'$ of operations initiated after $t$.
(As sites do not have access to a common clock, these sets cannot be
computed, but they exist nonetheless.)

Any operation $o \in O$ is either concurrent or happens-before any
operation $o' \in O'$.
If ${o} \concurrent {o'}$, then $\sequence{o \seq \ldots \seq o'} \equiv
\sequence{o' \seq \ldots \seq o}$; we need consider only the former.
If ${o} \happensbefore {o'}$, then the only legal order is \sequence{o
\seq \ldots \seq o'}.
The operations in $O$ are eventually executed at all sites, and the
operations in $O'$ execute after.
Thus the state at all sites has as common prefix the operations in $O$.

\bibliographystyle{plain}
\bibliography{bib,extrabib,predef}
\end{document}